\documentclass[12pt]{article}

\usepackage{amssymb,amsmath,amsthm,epsfig}

\theoremstyle{plain}
\newtheorem{lem}{Lemma}
\newtheorem{thm}[lem]{Theorem}

\newtheorem{pro}[lem]{Proposition}

\theoremstyle{definition}

\theoremstyle{remark}
\newtheorem{rem}{Remark}

\newcommand{\n}{\noindent}

\begin{document}

\title{The Universality of the Quantum Fourier Transform\\
in Forming the Basis of Quantum Computing Algorithms}

\author{Charles M.~Bowden$^1$, Goong Chen$^{2,3}$, Zijian Diao$^2$ and Andreas
Klappenecker$^{4}$}

\date{}
\maketitle

\begin{center}{\bf ABSTRACT}
\end{center}\smallskip

The quantum Fourier transform (QFT) is a powerful tool in quantum computing.
The main ingredients of QFT are formed by the Walsh-Hadamard transform $H$ and
phase shifts $P(\cdot)$, both of which are $2\times 2$ unitary matrices as
operators on the two-dimensional 1-qubit space. In this paper, we show that
$H$ and $P(\cdot)$ suffice to generate the unitary group $U(2)$ and,
consequently, through controlled-$U$ operations and their concatenations, the
entire unitary group $U(2^n)$ on $n$-qubits can be generated. Since any
quantum computing algorithm in an $n$-qubit quantum computer is based on
operations by  matrices in $U(2^n)$, in this sense we have the universality of
the QFT.
\vfil

\begin{itemize}
\item[(1)] U.S.~Army Missile Command, RD\&E Center, Redstone Arsenal, AL \
35898-5248.\newline E-mail: \ cmbowden@ro.com.
\item[(2)] Department of Mathematics, Texas A\&M University, College Station,
TX \ 77843-3368.\newline
E-mails: \ gchen@math.tamu.edu, zijian.diao@math.tamu.edu.
\item[(3)] Supported in part by Texas A\&M University Interdisciplinary
Research Initiative IRI 99-22.
\item[(4)] Institute for Algorithms and Cognitive Systems, Universit\" at
Karlsruhe, D-76128 Karlsruhe, Germany. E-mail: \ klappi@ira.uka.de.
\end{itemize}
\newpage

\section{Introduction}\label{sec1}

\indent

The quantum Fourier transform (QFT) on the additive group of integers modulo
$2^m$ is defined by
\begin{equation}\label{eq1}
\mathcal{F}_{2^m}(|a\rangle) = \sum^{2^m-1}_{y=0} e^{(2\pi iay)/2^m}
|y\rangle, \quad \text{for}\quad a\in \{0,1,2,\ldots, 2^m-1\}.
\end{equation}
QFT plays a significant  role in the development of the quantum
computer (QC). One may note, for example, that the potentially powerful 
integer factoring algorithm by P.~Shor relies critically on the QFT
for the detection of periodicity springing from the prime factors.

We can further analyze (\ref{eq1}) as follows. First, write 
\begin{align*}
a &= a_12^{m-1} +
a_2 2^{m-2} +\cdots+ a_{m-1}2^1 +a_m2^0 = (a_1a_2\ldots a_m)\\
\intertext{and}
y &= y_12^{m-1} + y_22^{m-2} +\cdots+ y_{m-1}2^1 +y_m 2^0 = (y_1y_2\ldots y_m).
\end{align*}
Then it is well known that
\begin{align}
\text{RHS of (\ref{eq1})} &= \sum^{2^{m-1}}_{y=0} e^{(2\pi iay/2^m)}|y_1
\ldots y_m\rangle\nonumber\\
&= \sum^{2^{m-1}}_{y=0} e^{2\pi i(0.a_m)y_1} |y_1\rangle e^{2\pi
i(0.a_{m-1}a_m)y_2} |y_2\rangle \cdots e^{2\pi i(0.a_1a_2\ldots a_m)y_m}
|y_m\rangle\nonumber\\
\label{eq2}
&= (|0\rangle + e^{2\pi i(0.a_m)} |1\rangle) (|0\rangle + e^{2\pi i(0.a_{m-1}
a_m)} |1\rangle) \cdots (|0\rangle + e^{2\pi i(0.a_1a_2\ldots a_m)}|1).
\end{align}
In the above factorization (or ``untangling''), each factor is of the form
\begin{equation}\label{eq3}
|0\rangle + e^{i\omega} |1\rangle.
\end{equation}
Such a state can be produced in two steps [\ref{Cl}, pp.~340--341]: \ First,
apply the transformation
\begin{equation}\label{eq4}
H = \frac1{\sqrt 2} \left[\begin{array}{cr}
1&1\\ 1&-1\end{array}\right],
\end{equation}
where $H$ is known as the Walsh-Hadamard transform, to the state $|0\rangle$:
\begin{equation}\label{eq5}
H |0\rangle = \frac1{\sqrt 2} (|0\rangle + |1\rangle).
\end{equation}
Next, apply the phase shift operator
\begin{equation}\label{eq6}
P(\omega) = \left[\begin{array}{cc}
1&0\\ 0&e^{i\omega}\end{array}\right]
\end{equation}
to (\ref{eq5}), yielding
\begin{equation}\label{eq7}
P(\omega) [H|0\rangle] = \frac1{\sqrt 2} (|0\rangle + e^{i\omega} |1\rangle).
\end{equation}
The RHS of (\ref{eq7}) is (\ref{eq3}) (apart from a normalization
coefficient). Therefore, we see that the constituents of the QFT are $H$ and
$P(\omega)$. From the quantum optics point of view, $H$ is realized by a
half-silvered mirror (beam splitter) and $P(\omega)$ represents a phase
shifter, as in a standard Mach-Zehnder interferometer ([\ref{Cl}, \ref{Kl}]).

First, we wish to emphasize that the QFT strictly by itself is {\em not
universal\/} in quantum computing; see Remark~\ref{rem2} below. Thus, the
question becomes whether the two constituents $H$ an $P(\cdot)$ of QFT are
universal or not. The conjecture we want to pose here is the following:
\begin{align} \text{[Q]} &~~\text{``Any QC algorithm can be represented as a
composition of}\nonumber\\ \label{eq8}
&~~\text{Walsh-Hadamard transforms and associated conditional phase
shifts.''}
\end{align}

The implication of (\ref{eq8}) is that the realization of any QC algorithm
translates into a combination of elementary quantum interferometric
operations, i.e., single particle beam splitter (Walsh-Hadamard transform)
followed by a conditional phase shift. Any QC algorithm can thus be
formulated, or reformulated, in terms of elementary multiparticle quantum
interferometric operations. The unique universal fundamental properties of QC
concerning quantum superposition, entanglement and interference are all
explicitly represented in terms of quantum multiparticle interferometry (QMI).

QMI practically is not to be taken as a proposed embodiment of a QC any more
than the Turing machine is to be taken as a literal construction in classical
computing. Rather, Ekert [\ref{Ek}] has suggested its equivalence to QC in the
sense of its universality, meaning that QMI could be viewed as the closest QC
analogue of the classical Turing machine (through the universality theorem
established in this paper). This concept and viewpoint should provide physical
insights into the operational aspects and can facilitate efficient design of
a universal QC.

\section{Mathematical Proof of the Universality of $\pmb{H}$ and
$\pmb{P(\cdot)}$}\label{sec2}

\indent

Our answer to [Q] is affirmative. We now proceed to provide the mathematical
justifications below. 

As usual, we let $U(n)$ to denote the unitary group on $n$-dimensional space.
By abuse of notation, we regard $U(n)$ the same as the multiplicative group of
all $n\times n$ unitary matrices. $SO(n)$ denotes the orthogonal group on
$n$-dimensional spaces or, equally, the multiplicative group of all $n\times
n$ orthogonal matrices. We also define the {\em maximal torus\/} $T(n)$ in
$U(n)$ as
$$T(n) = \{\text{diag}(e^{i\omega_1},\ldots, e^{i\omega_n})\mid\omega_1,
\omega_2,\ldots, \omega_n\in \mathbb{R}\},$$
i.e., $T(n)$ consists of all $n\times n$ diagonal matrices whose diagonal
entries are complex numbers of unit magnitude. $T(n)$ is a subgroup of the
multiplicative group $U(n)$.

Let $\mathcal{A}$ be a collection of $n\times n$ unitary matrices. 
In this paper,
we will use $\mathcal{G}_n(\mathcal{A})$ to denote {\em the unitary subgroup
of $U(n)$ generated by\/} $\mathcal{A}$, i.e.,
$$\mathcal{G}_n(\mathcal{A}) = \bigcap_\alpha\{G_\alpha\mid G_\alpha \text{ is
a subgroup of } U(n), \mathcal{A} \subseteq G_\alpha\}.$$     
We will write $\mathcal{G}_n(\mathcal{A})$ simply as
$\mathcal{G}(\mathcal{A})$ if the value of $n$ is clear from the context. 

We begin with $n=2$.

\begin{lem}[$\lbrack$\ref{Ba}, Lemma~4.1$\rbrack$]\label{lem1}
We have $U(2) = \mathcal{G}(SO(2), T(2))$, i.e., $U(2)$ is generated by $SO(2)$
and $T(2)$; more precisely, for every $A\in U(2)$, we have
$$A = \left[\begin{array}{cc} e^{i\delta}&0\\ 0&e^{i\delta}\end{array}\right]
\left[\begin{array}{cc} e^{i\alpha/2}&0\\ 0&e^{-i\alpha/2}\end{array}\right]
\left[\begin{array}{rc} \cos \omega&\sin\omega\\ -\sin\omega&\cos\omega
\end{array}\right] \left[\begin{array}{cc} e^{i\beta/2}&0\\ 0&e^{-i\beta/2}
\end{array}\right],$$
for some $\alpha,\beta,\delta,\omega\in \mathbb{R}$.$\hfill\square$
\end{lem}

\begin{lem}\label{lem2}
$T(2) \subseteq \mathcal{G}(H,P(\cdot))$.
\end{lem}

\begin{proof}
We first note that the NOT-gate
\begin{equation}\label{eq7a}
X \equiv \left[\begin{array}{cc}
0&1\\ 1&0\end{array}\right]
\end{equation}
can be obtained as
\begin{equation}\label{eq8a}
X = HP(-\pi)H.
\end{equation}
Therefore $X\in \mathcal{G}(H,P(\cdot))$. From this, we have 
\begin{equation}\label{eq9}
XP(\omega_1) XP(\omega_2) = \left[\begin{array}{cc}
0&1\\ 1&0\end{array}\right] \left[\begin{array}{cc}
1&0\\ 0&e^{i\omega_1}\end{array}\right] \left[\begin{array}{cc}
0&1\\ 1&0\end{array}\right] \left[\begin{array}{cc}
1&0\\ 0&e^{i\omega_2}\end{array}\right] = \left[\begin{array}{cc}
e^{i\omega_1}&0\\ 0&e^{i\omega_2}\end{array}\right],
\end{equation}
for any given $\omega_1,\omega_2\in \mathbb{R}$. Therefore $\mathcal{G}(H,
P(\cdot))$ contains the maximal torus $T(2)$.
\end{proof}

\begin{lem}\label{lem3}
$SO(2) \subseteq \mathcal{G}(H, P(\cdot))$.
\end{lem}

\begin{proof}
For each rotation matrix
$$R(\omega) = \left[\begin{array}{rl}
\cos \omega&\sin\omega\\ -\sin\omega&\cos\omega\end{array}\right],$$
we easily verify that
\begin{equation}\label{eq10}
R(\omega) = P\left(\frac\pi2\right) HP(\omega)X P(-\omega) HP\left(-
\frac\pi2\right).
\end{equation}
\end{proof}

\begin{thm}\label{thm4}
$\mathcal{G}(H,P(\cdot)) = U(2)$.
\end{thm}

\begin{proof}
This follows immediately from Lemmas~\ref{lem1}--\ref{lem3}.
\end{proof}

At this point, it should already be clear from the results in [\ref{Ba}] that
$U(2^n)$ {\em can be generated through controlled-$U(2)$ gates}, for any
$n=1,2,\ldots$. To make this paper sufficiently self-contained, however, let
us give the following concise, rigorous treatment as to how to construct any
$V\in U(2^n)$ from a serial connection of a collection of unitary matrices
$V_{ij}$, where each $V_{ij}$ is a (generalized) controlled-$U(2)$ gate. The
precise statement is given below.

\begin{thm}\label{thm5}
Let $V\in U(2^n)$. Then
\begin{equation}\label{eq11}
V  = \prod^{2^n-1}_{i=1} \prod^{i-1}_{j=0} V_{ij}.
\end{equation}
for a collection of matrices $V_{ij}\in U(2^n)$ such that
\begin{equation}\label{eq12}
\left\{\begin{array}{l}
V_{ij}\colon \ \mathcal{S}_{ij} \longrightarrow \mathcal{S}_{ij} \text{ is the
identity transformation,}\\
\mathcal{S}_{ij} \equiv \text{span}\{|m\rangle\mid m\in \{0,1,\ldots, 2^n-1\},
m\ne i, m\ne j\},\\
\quad 0 \le j < i \le 2^n-1.\end{array}\right\}
\end{equation}
In other words, each $V\in U(2^n)$ is a product of (generalized)
controlled-$U(2)$ unitary matrices $V_{ij}$, which acts nontrivially only on
$\mathcal{S}^\bot_{ij} = \text{span}\{|i\rangle, |j\rangle\}$.
\end{thm}

\begin{proof}
We first quote the following fact [\ref{Mu}, \ref{Re}]: \ For any $V\in
U(2^n)$, there exists a collection of unitary matrices $T_{i,j}$, $0 \le j < i
\le 2^n-1$, and a $D\in T(2^n)$ such that
\begin{equation}\label{eq12a}
V = \left(\prod^{2^n -1}_{i=1} \prod^{i-1}_{j=0} T_{i,j}\right)D,
\end{equation}
where $T_{i,j}\in SO(2^n) \subseteq U(2^n)$ is a rotation involving
$|i\rangle$ and $|j\rangle$ and satisfying (\ref{eq12}). For the benefit of
the reader and for the sake of self-containedness, we include a direct proof
of (\ref{eq12a}) in the Appendix, condensed from [\ref{Mu}].

Now we can break up $D$ into
\begin{equation}\label{eq13}
D=
\left( \begin{array}{cccc}
d_0 &     &   &  \\
    & d_1 &   &  \\
    &     &\ddots &  \\
    &     &       & d_{2^n-1}
\end{array} \right)  
= D_1 D_2 \ldots D_{2^n-1}
\end{equation}
where  
\begin{equation}\label{eq14}
D_1=
\left( \begin{array}{ccccc}
d_0 &   0 &    &  & \\
 0  & d_1 &    &  &  \\
    &     & 1 &  &  \\
    &     &   & \ddots &  \\
    &     &   &        & 1
\end{array} \right)  
 \end{equation}
and 
\begin{equation}\label{eq15}
D_i=
\left( \begin{array}{ccccc}
1 &     &   &   & \\
    & \ddots &   &   & \\
    &     &d_i &  &  \\
    &     &       & \ddots & \\
    &     &    &     &1
\end{array} \right)  
 \end{equation}
for $i=2, 3, \ldots, 2^n-1$.  It is easy to see that 
$D_1$ acts trivially except on $|0\rangle $ and $|1\rangle $, and the other
$D_i$'s act non-trivially only on $|i\rangle $.  In addition, $D_i$'s commute
with each other, and each $D_i$ commutes with $T_{k,l}$, 
$\forall 0\le l < k <i$ as well.  Thus,
\begin{align}
V = &~T_{2^n-1,2^n-2}\ldots T_{2^n-1,0} T_{2^n-2,2^n-3}\ldots T_{2^n-2,0}
\ldots T_{2,1}T_{2,0}T_{1,0} D_1 D_2 \ldots D_{2^n-1}\nonumber\\
\label{eq16}
 &\!\!\!\!\!\!\!\!\left.\begin{array}{l}
= T_{2^n-1,2^n-2}T_{2^n-2,2^n-3}\ldots T_{2^n-1,0}D_{2^n-1}\\ 
~~~T_{2^n-2,2^n-3}\ldots T_{2^n-2,0}D_{2^n-2}\\
~~~\ldots \ldots \\
~~~T_{2,1}T_{2,0} D_2 \\
~~~T_{1,0}  D_1\end{array}\right\} \ 2^n-1 \text{ strings of products}
\end{align}

For $0\le j < i \le 2^n-1$, define 
\[
V_{ij}=\left\{ \begin{array}{ll}
T_{i,j} &  \textrm{if $j \ne 0$,}\\
T_{i,j}D_i = T_{i,0}D_i & \textrm{if $j=0$.}
\end{array} \right.
\]
Therefore we have reached
\[ 
V=\prod _{i=1}^{2^n-1}\prod _{j=0}^{i-1} V_{ij}
\]
where each $V_{ij}$ is a unitary matrix which acts nontrivially only on the
states $|i\rangle $ and $|j\rangle$ satisfying (\ref{eq12}).
\end{proof}

\begin{rem}\label{rem1}
\begin{itemize}
\item[(1)] In Barenco et al.\ [\ref{Ba}, p.~3465, right column, line 34], the
equation there corresponds to our equation (\ref{eq12a}) here. However, a
summation sign $\sum$ is used instead of the product sign $\prod$ (which is
actually a double product $\prod\limits_i \prod\limits_j$ in our
(\ref{eq12a})) which, of course, is a misprint.
\item[(2)] The factoring of $D$ in (\ref{eq13}) into the product of
$D_1,D_2,\ldots$ and $D_{2^n-1}$ in the form of (\ref{eq14}) and (\ref{eq15})
is peculiar in the sense that $D_1$ is chosen differently from the other
$D_i$'s, $i\ne 1$. It must be done this way (but no further mathematical
explanations were given in [\ref{Ba}]). The reason for this is that there are
$2^n-1$ strings of products as indicated in (\ref{eq16}). Therefore $D$ must be
factorized to have $2^n-1$ factors $D_1,D_2,\ldots, D_{2^n-1}$, in the unique
way of (\ref{eq14}) and (\ref{eq15}) in order to satisfy (\ref{eq12}).
$\hfill\square$
\end{itemize}
\end{rem}

\begin{rem}\label{rem2}
Now it can be readily seen that the QFT itself is not universal in the sense
that $U(2^n)$ is not generated by $\mathcal{F}_{2^n}$ (cf.\ (\ref{eq1}), with
$m=n$ therein) or (generalized) controlled-$\mathcal{F}_{2^m}$ (where $m<n$)
operations. First, check $n=1$: \ we see that $\mathcal{F}_{2^n} =
\mathcal{F}_2$ is actually the Walsh-Hadamard transform $H$ (apart from the
normalization factor $1/\sqrt 2$). Therefore, the phase shifts $P(\omega)$ in
(\ref{eq6}) cannot be generated by $\mathcal{F}_2$ because $P(\omega)$ has
eignevalues 1 and $e^{i\omega}$ while $H$ has eigenvalues 1 and $-1$. For a
general positive integer $n$, the range of $\mathcal{F}_{2^n}$ or of
controlled-$\mathcal{F}_{2^m}$, $m<n$, consists at most of linear combinations
of states of the form
$$e^{2\pi i[(0.a_n)y_1+(0.a_{n-1}a_n)y_2 +\cdots+ (0.a_1\ldots a_n)y_n]} |y_1
\ldots y_n\rangle, \text{ where } a_j,y_j\in \{0,1\}, \text{ for }
j=1,2,\ldots, n.$$
The phases of such states are {\em not even dense\/} with respect to all
possible phases $e^{2\pi i\theta}$, $0\le \theta <2\pi$.$\hfill\square$
\end{rem}

\section{Remarks on Circuits}\label{sec3}

\indent

The decomposition (\ref{eq11}) is a mathematical rendering of statement [Q] and
answers the conjecture affirmatively. In this section, let us further
elaborate on the circuit design aspects, based on the work in Barenco et al.\
[\ref{Ba}, \S VIII] and [\ref{Kl}].

Each factor $V_{ij}$ in (\ref{eq11}) satisfies (\ref{eq12}) and thus $V_{ij}$
acts nontrivially only on the states $|i\rangle$ and $|j\rangle$. Denote the
restriction of $V_{ij}$ to the 2-dimensional subspace $\mathcal{S}^\bot_{ij} =
\text{span}\{|i\rangle, |j\rangle\}$ by $\widehat V_{ij}$. Then $\widehat
V_{ij} \in U(2)$. As pointed out in [\ref{Ba}, p.~3465], each $V_{ij}$ is not
a standard $\Lambda_{n-1}(\widehat V_{ij})$ (in the notation of [\ref{Ba},
p.~3458]) gate in the sense that the {\em controls are states rather than
bits}.

Nevertheless, using Proposition~\ref{pro9} below, Barenco et al.\ [\ref{Ba},
\S VIII] point out how to rearrange basis states with a ``gray code connecting
state $|i\rangle$ to state $|j\rangle$'' such that $V_{ij}$ becomes unitarily
equivalent to $\Lambda_{n-1}(\widehat V_{ij})$.  In this sense, $V_{ij}$ are
{\em generalized\/} controlled-$\widehat V_{ij}$ gates.

\begin{pro}\label{pro9}
The symmetric group $S_{2^n}$ of permutations on the symbols $0,1,2,\ldots,
2^n-1$ is generated by the 2-cycle $(2^n-2, 2^n-1)$ and the $2^n$-cycle
$(0,1,2,\ldots, 2^n-1)$.
\end{pro}

\begin{proof}
This is a basic fact which can be found in most basic algebra or group theory
books.
\end{proof}

Incidentally, we note that the 2-cycle $(2^n-2, 2^n-1)$ is a permutation
between the states $|\underbrace{1~1 \ldots 1~0}_{n \text{ bits}}\rangle$ and
$|\underbrace{1~1 \ldots 1}_{n \text{ bits}}\rangle$ and thus can be realized
by the controlled-NOT gate with the $n$th qubit as the {\em target bit\/} and
the first $(n-1)$ bits as the {\em control bits\/} as shown in Fig.~1.

On the other hand, the $2^n$-cycle $(0,1,2,\ldots, 2^n-1)$ makes the rotation
of the states $|0\rangle \to |1\rangle \to\cdots\to |2^n-2\rangle \to 
|2^n-1\rangle \to |0\rangle$, i.e., the $|x\rangle \to |x+1 \bmod 2^n\rangle$
operation. This can be implemented by the circuit as shown in Fig.~2.

Therefore, any permutation of the basis states $|x\rangle$, $x=0,1,2,\ldots,
2^n-1$, can be realized by finitely many controlled-NOT operations consisting
of circuits as shown in Figs~1 and 2.

Thus, each factor $V_{ij}$ in (\ref{eq11}) can be realized by the circuit as
shown in Fig.~3.
\medskip

\begin{center}
\epsfig{figure=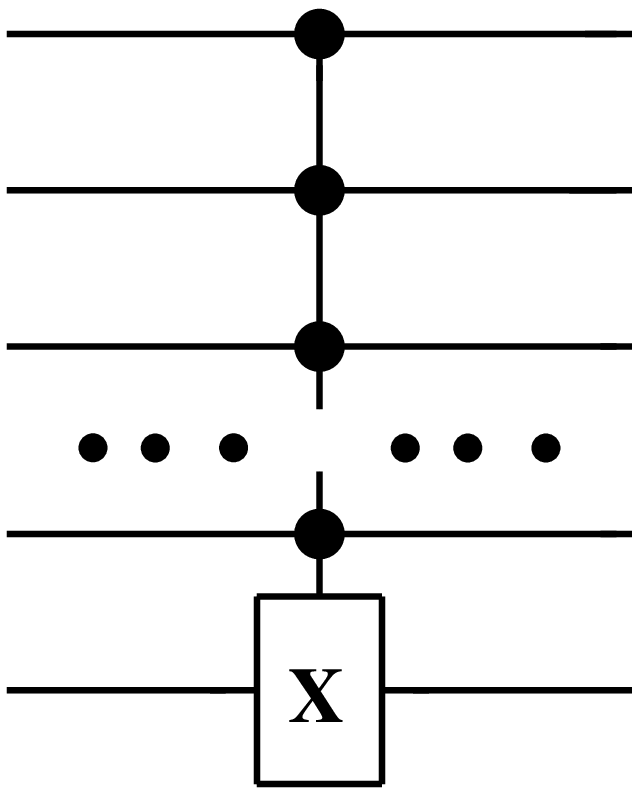,height=2in,width=1.7in}
\end{center}

\n {\bf Fig.~1~~The $\pmb{n}$-bit controlled-NOT gate
$\pmb{\Lambda_{n-1}(X)}$, where $\pmb{X}$ is given by (\ref{eq7a}). This gate
implements the two cycle $\pmb{(2^n-2, 2^n-1)}$ in Proposition~\ref{pro9}.}

\begin{center}
\epsfig{figure=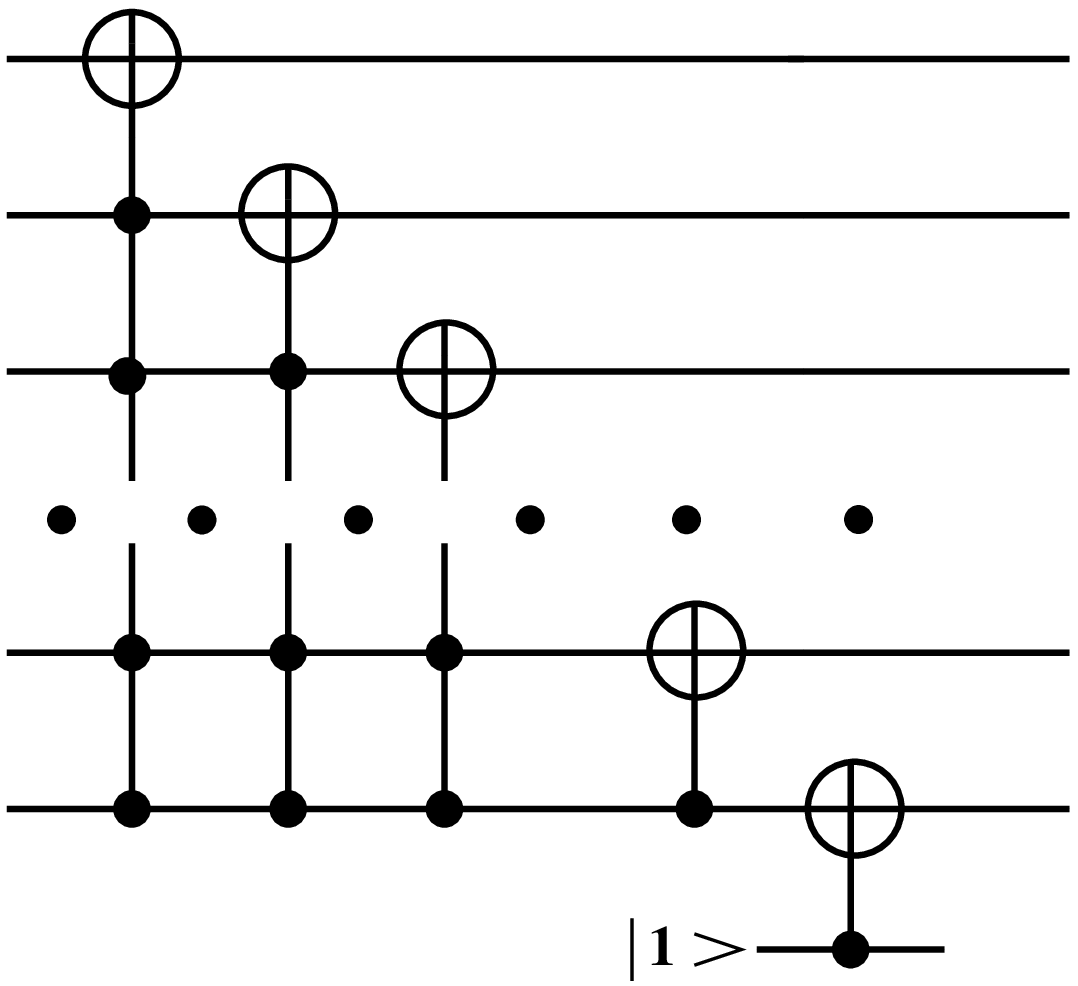,height=2.7in,width=3in}
\end{center}

\n {\bf Fig.~2~~This circuit implements the operation $\pmb{|x\rangle\to |x+1
\bmod 2^n\rangle}$ or, equivalently, the $\pmb{2^n}$-cycle $\pmb{(0,1,2,\ldots,
2^n-1)}$ in Proposition~\ref{pro9}. Note that the bit $\pmb{|1\rangle}$ at
the bottom of the figure is the ``scratch bit'' which is sometimes omitted in
circuit drawing. All the gates in this circuit are controlled-NOT gates.}
\bigskip
\newpage

\begin{center}
\epsfig{figure=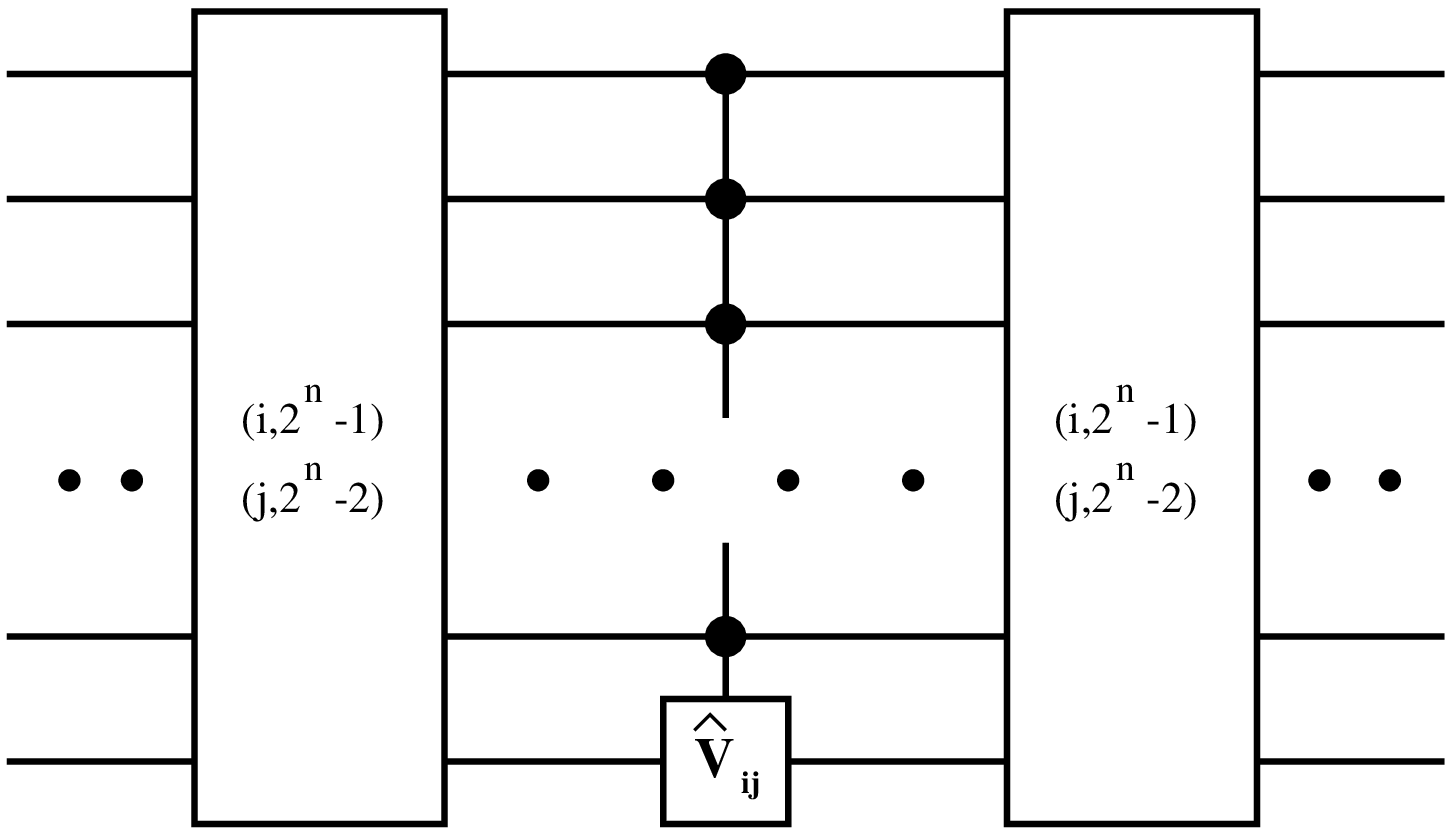,height=2.2in,width=4in}
\end{center}

\n {\bf Fig.~3~~The unitary matrix $\pmb{V_{ij}}$ in (\ref{eq11}) as a
controlled-$\pmb{\widehat V_{ij}}$ gate where $\pmb{\widehat V_{ij}\in
U(2)}$. The operations $\pmb{(i,2^n-1)}$ and $\pmb{(j,2^n-2)}$ in the two
boxes are cyclic permutations (which can be realized by concatenations of
circuits in Figs.~1 and 2).} \bigskip

By concatenating together all the blocks $V_{ij}$ as shown in Fig.~3 according
to the factorization (\ref{eq11}), we have constructed all $V\in U(2^n)$ with
controlled-$\widehat V_{ij}$ gates according to (\ref{eq11}). Each $\widehat
V_{ij}\in U(2)$ is then further formed from concatenations of the gates
$H,P(\omega) \in U(2)$ by Theorem~\ref{thm4}. It is in this sense that we have
the universality of the Walsh-Hadamard gate $H$ and the phase shift gate
$P(\cdot)$ and, consequently, that of the quantum Fourier transform with the
affirmative answer to question [Q] in (\ref{eq8a}).

\section*{Appendix: \ Decomposition Procedure of General Finite Dimensional
Unitary Transformations into a Product of Plane Unitary Transformations}

\indent

First, we define a special type of unitary transformations
$T_{pq}(\phi,\sigma) \in U(n)$ by
$$T_{pq}(\phi,\sigma) = [t_{ij}]_{n\times n},\qquad 1\le  p,q\le n,\quad p\ne
q,$$
where
$$t_{ij} = \left\{\begin{array}{cl}
1,&i=j, i\ne p, i\ne q,\\
\cos \phi,&i=j=p \text{ or } i=j=q,\\
0,&i\ne j, i\ne p, j\ne q \text{ and } i\ne q, j\ne p,\\
-e^{-i\sigma}\sin\phi,&i=p \text{ and } j=q,\\
e^{i\sigma} \sin\phi,&i=q \text{ and } j=p;
\end{array}\right.$$
i.e.,
$$T_{pq}(\phi,\sigma) = \begin{matrix}
p~~~~~~~~~~~q\\
\begin{matrix}
\\
p\\
q\end{matrix}
\left[\begin{matrix}
1&0&0&&&&&&0\\
0&1\\
&&1\\
&&&\ddots\\
&&&&\cos \phi&-e^{-i\sigma}\sin \phi\\
&&&&e^{i\sigma}\sin\phi&\cos\phi\\
&&&&&&1&&0\\
0&0&&&&&&\ddots\\
&&&&&&&0&1\end{matrix}\right].\end{matrix}$$
$T_{pq}(\phi,\sigma)$ is just a plane unitary transformation acting
non-trivially only on states  $p$ and $q$.

Let $V\in U(n)$. We want to find some $T_{n,n-1}(\phi,\sigma)$ such that
$T^*_{n,n-1}V = V' = [v'_{ij}]_{n\times n}$, where $v'_{n-1,n} = 0$:
$$T^*_{n,n-1}V = \left[\begin{matrix}
1&0&0\\
0&1&0\\
0&0&1&&~~~~~~~~~~~~~~\bigcirc\\
&&&\ddots\\
&&&&\cos\phi&e^{-i\sigma}\sin \phi\\
&&\bigcirc&&-e^{i\sigma}\sin\phi&\cos\phi\end{matrix}\right]
\left[\begin{matrix}
v_{11}&\ldots&v_{1,n-1}&v_{1n}\\
\vdots&&\vdots&\vdots\\
v_{n-1,1}&\ldots&v_{n-1,n-1}&v_{n-1,n}\\
v_{n1}&\ldots&v_{n,n-1}&v_{nn}\end{matrix}\right],$$
so
$$v'_{n-1,n} = v_{n-1,n} \cos \phi + v_{nn} e^{-i\sigma} \sin \phi.$$
We consider all possibilites:
\begin{itemize}
\item[Case 1:] $v_{n-1,n} = 0$. Then we choose $\phi=0, \sigma=0$, i.e.,
$T_{n-1,n}(\phi,\sigma) = I_n$, and we obtain $v'_{n-1,n} = v_{n-1,n} =
0$.
\item[Case 2:] $v_{n-1,n}\ne 0, v_{nn}=0$. Then choose $\phi = \pi/2,
\sigma=0$. Obtain $v'_{n-1,n}=0$.
\item[Case 3:] $v_{n-1,n} \ne 0, v_{nn}\ne 0$. Write $v_{n-1,n} =
r_{n-1,n}e^{i\theta_{n-1,n}}$, $v_{nn} = r_{nn} e^{i\theta_{nn}}$.
Choose $\sigma = -\theta_{n-1,n} + \theta_{nn}$ and $\phi = \tan^{-1}
(-r_{n-1,n}/r_{nn})$. Obtain
\begin{align*}
v'_{n-1,n} &= \cos \phi\cdot r_{n-1,n} e^{i\theta_{n-1,n}} + \sin
\phi\cdot r_{nn} e^{i(-\sigma+\theta_{nn})}\\
&= \left(\frac{r_{n-1,n}}{r_{nn}} + \tan\phi\right) r_{nn}\cos \phi
e^{i\theta_{n-1,n}} = 0.
\end{align*}
\end{itemize}
Therefore, we have found $T_{n,n-1} \in U(n)$ such that
$$T^*_{n,n-1}V = \left[\begin{array}{cc}
\begin{array}{ccccc|}
\noalign{\vspace{-9pt}}
*&&\ldots&&*\\
\vdots&&&&\vdots\\
*&&\ldots&&*\\
\multispan5\hrulefill\end{array}&\begin{array}{c}
\noalign{\vspace{-6pt}}
{v'_{1n}}\\
\vdots\\
\noalign{\vspace{-5pt}}
v'_{n-2,n}\\
\noalign{\vspace{-4pt}}
0\end{array}\\
~~v'_{n1}\quad \ldots\quad v'_{n,n-1}&v'_{nn}\end{array}\right].$$
Similarly, we can find $T_{n,n-2}, T_{n,n-3},\ldots,T_{n,1}$ such that
\begin{align*}
T^*_{n,n-2} &T^*_{n,n-1}V = \left[\begin{matrix}
*&&*&v''_{1n}\\
\vdots&&\vdots&\vdots\\
\vdots&&\vdots&v''_{n-3,n}\\
&&&0\\
*&&*&0\\
v''_{n1}&\ldots&v''_{n,n-1}&v''_{nn}\end{matrix}\right],\\
&\vdots\\
T^*_{n1} T^*_{n2} &\ldots T^*_{n,n-2} T^*_{n,n-1}V = \left[\begin{matrix}
*&&*&0\\
\vdots&&\vdots&0\\
\vdots&&\vdots&\vdots\\
*&&*&0\\
\tilde v_{n1}&\ldots&\tilde v_{n,n-1}&\tilde v_{nn}\end{matrix}\right] \equiv
W. \end{align*}
Since $W$ is unitary, we conclude $\tilde v_{n1} = \tilde v_{n2} = \cdots =
\tilde v_{n,n-1} = 0$ and $\tilde v_{nn} = e^{i\alpha_n} \equiv d_n$ for some
$\alpha_n\in \mathbb{R}$. Thus
$$T^*_{n1}T^*_{n2}\ldots T^*_{n,n-2} T^*_{n,n-1}V = \left[\begin{array}{cc}
\begin{array}{ccc|}
&&\\
&**&\\
&&\\
\multispan3\hrulefill\end{array}&\begin{array}{c}
0\\ \vdots\\ 0\end{array}\\
0\quad \ldots\quad 0&d_n\end{array}\right].$$
Now, applying  the same technique to the remaining $(n-1)\times (n-1)$
undiagonalized matrix block $(**)$ above, together with a simple induction
argument, we obtain plane unitary transformation $T_{n1},\ldots, T_{n,n-1},
T_{n-1,1}, \ldots, T_{n-1,n-2},\ldots, T_{31}, T_{32}$ and $T_{21}$ such that
$$T^*_{21}T^*_{31}T^*_{32}T^*_{41}\ldots T^*_{n-1,1}\ldots T^*_{n-1,n-2}
T^*_{n1} \ldots T^*_{n,n-1}V = \left[\begin{matrix}
d_1\\
&d_2&&0\\
&&\ddots\\
&0&&d_n\end{matrix}\right]=D,$$
 where $d_j = e^{i\alpha_j}$ for $j=1,2,\ldots, n$.

Therefore
\begin{align*}
V &= T_{n,n-1}, T_{n,n-2}\ldots T_{n1} T_{n-1,n-2}\ldots T_{n-1,1} \ldots
T_{32} T_{31} T_{21}D\\
&= \left(\prod^n_{i=1} \prod^{i-1}_{j=1} T_{i,j}\right)D
\end{align*}
and (\ref{eq12a}) is proved.

   \newpage

\section*{References}

\begin{enumerate}
\item\label{Ba}
A. Barenco, C.H. Bennett, R. Cleve, D.P. Divincenzo, N. Margolus, P. Shor, 
T. Sleator, J.A. Smolin, and H. Weinfurter,
Elementary gates for quantum computation, 
Physical Review A, 52(5) (1995), 3457--3467.

\item\label{Cl}
R.\ Cleve, A. Ekert, C. Macchiavello and M. Mosca, Quantum algorithms
revisited, Proc.\ R.\ Soc.\ London A 454 (1998), 339--354.

\item\label{Ek}
A. Ekert, Quantum interferometers as quantum computers, Phys.\ Scripta T76
(1998), 218--222.

\item\label{Kl}
A. Klappenecker, 
Computing with a quantum flavor, manuscript (2000).

\item\label{Mu}
F.D. Murnaghan, {\em The Unitary and Rotation Groups}, Spartan, Washington,
D.C., 1962.

\item\label{Re}
M. Reck, A. Zeilinger, H.J. Bernstein, and P. Bertani, 
Experimental realization of any discrete unitary operator, 
Physical Review Letters  73 (1994), 58--61.
\end{enumerate}

\end{document}